\relax
\documentclass[letterpaper]{article} 
\usepackage{aaai21}  
\usepackage{times}  
\usepackage{helvet} 
\usepackage{courier}  
\usepackage[hyphens]{url}  
\usepackage{graphicx} 
\usepackage[square,numbers,sort&compress]{natbib}  
\usepackage{caption} 
\usepackage{epsfig}
\usepackage{subcaption}
\usepackage{amsmath}
\usepackage[usenames,dvipsnames,svgnames,table]{xcolor}
\usepackage{amssymb}
\usepackage{setspace}
\usepackage{times}
\usepackage{amsthm}
\usepackage{hyperref} 
\usepackage{booktabs}
\usepackage{multirow}
\usepackage{tabularx}
\usepackage{threeparttable}
\usepackage[graphicx]{realboxes}

\hypersetup{unicode=false, pdftoolbar=true, pdfmenubar=true, pdffitwindow=false, pdfstartview={FitH}, pdfcreator={Daniel Larremore}, pdfproducer={Daniel Larremore}, pdfkeywords={} {} {}, pdfnewwindow=true, colorlinks=true, linkcolor=red, citecolor=Green, filecolor=magenta, urlcolor=cyan,}
\graphicspath{{../../../../figures/}} 

\newcommand{\beginsupplement}{%
        \setcounter{table}{0}
        \renewcommand{\thetable}{S\arabic{table}}%
        \setcounter{figure}{0}
        \renewcommand{\thefigure}{S\arabic{figure}}%
     }

\DeclareCaptionFormat{subfont}{\fontsize{3}{4}\selectfont#1#2#3}

\title{An Open-Source Cultural Consensus Approach to Name-Based Gender Classification}
\author {
    Ian Van Buskirk,\textsuperscript{\rm 1}
    Aaron Clauset, \textsuperscript{\rm 1, \rm 2, \rm 3}
    Daniel B. Larremore \textsuperscript{\rm 1, \rm 2} \\
}
\affiliations {
    \textsuperscript{\rm 1} Department of Computer Science, University of Colorado Boulder \\
    \textsuperscript{\rm 2} BioFrontiers Institute, University of Colorado Boulder \\
    \textsuperscript{\rm 3} Santa Fe Institute \\
    ian@colorado.edu, aaron.clauset@colorado.edu, daniel.larremore@colorado.edu
}

\begin{document}

\maketitle

\begin{abstract}
Name-based gender classification has enabled hundreds of otherwise infeasible scientific studies of gender.
Yet, the lack of standardization, proliferation of ad hoc methods, reliance on paid services, understudied limitations, and conceptual debates cast a shadow over many applications.
To address these problems we develop and evaluate an ensemble-based open-source method built on publicly available data of empirical name-gender associations.
Our method integrates 36 distinct sources—spanning over 150 countries and more than a century—via a meta-learning algorithm inspired by Cultural Consensus Theory (CCT). We also construct a taxonomy with which names themselves can be classified.
We find that our method's performance is competitive with paid services and that our method, and others, approach the upper limits of performance; we show that conditioning estimates on additional metadata (e.g. cultural context), further combining methods, or collecting additional name-gender association data is unlikely to meaningfully improve performance.
This work definitively shows that name-based gender classification can be a reliable part of scientific research and provides a pair of tools, a classification method and a taxonomy of names, that realize this potential.
\end{abstract}

\section{Introduction}

Name-based gender classification, the process of assigning individuals gendered labels based on their names alone, has been used in a diverse range of scientific studies, including investigations of gender's role in the bias of juries~\cite{flanagan2018race}, disparities in homeownership~\cite{shiffer2020gender}, the careers of US ambassadors~\cite{arias2018tenure}, participation in Alaskan fisheries~\cite{szymkowiak2020genderizing}, behavior on social media~\cite{smith2019mapping,jain2017location}, representation in encyclopedias~\cite{reagle2011gender}, and how the COVID-19 pandemic has disproportionately impacted women scientists~\cite{vincent2020decline,wehner2020comparison,squazzoni2020no}.
In studies of gender and academia, past work using name-based gender classification has investigated differing publication rates~\cite{fox2018patterns,thomas2019gender}, 
patterns of collaboration~\cite{ouyang2019association,bravo2019gender}, 
peer review~\cite{squazzoni2021peer,williams2018role}, 
citation practices~\cite{mohammad2020gender,dworkin2020extent}, and broader issues of representation~\cite{james2019gender,stathoulopoulos2019gender},
with sample sizes ranging from only a few thousand~\cite{nittrouer2018gender,chari2017gender,vallence2019data} 
to millions 
\cite{west2013role,lariviere2013bibliometrics,sugimoto2019factors,huang2020historical}.

Among past studies using name-based gender classification tools, the majority have relied upon paid services, which often provide an API that takes as input a name and returns either a gender-binary label or a label along with an uncertainty estimate. The use of such paid services is problematic in three distinct ways.
First, paid services lack transparency about their methods, data, and assumptions. As such, it is difficult to interpret their classifications and  scalar confidence scores, or audit and extend these services.
Second, paid services are oriented towards applications in industry, not towards academic research, resulting in an implicit conflict from studying topics like representation and bias with tools designed for targeted advertising that tends to exploit or reaffirm those biases.
Third, research using paid services would incur unnecessary costs, were free alternatives available.

It follows that there is a critical need for open-source, open-data and freely available methods which give researchers a greater degree of control over and understanding of their work. To meet this need, one frequently attempted solution is to use publicly available datasets of names and gender or sex associations, such as the U.S. Social Security Administration database~\cite{dworkin2020extent, west2013role, ouyang2019association, szymkowiak2020genderizing, mohammad2020gender, hofstra2020diversity, helmer2017gender, king2017men}, to establish empirical associations that enable name-based gender classification. 
Unfortunately, although a number of publicly available datasets exist and have been used in isolation, there is currently no open-source method whose performance---measured by the fraction of individuals whose name-based gender classifications match given labels collected via manual annotations or other means---is competitive with paid services~\cite{santamaria2018comparison, berube2020wiki}. 
Past attempts to use publicly available data, although creative, have made use of at most a few datasets and as a result could cause unforeseen issues for analyses (e.g. publicly available data from the United States could suggest a name is gendered female when in a global context the name is gendered male, as in the case of Jan)~\cite{dworkin2020extent, cheong2021computer}.
This risk has stimulated work that seeks to identify names that may pose challenges to classifiers~\cite{blevins2015jane,smith2013search,torvik2016ethnea}.
However, lacking a rich enough set of public data, our general understanding of name-gender associations remains underdeveloped, and much confusion remains around the severity of problems that are thought to arise when such associations are used to classify individuals.

To address these issues, we present a collection of publicly available name-gender association data in a standardized format, and use it to develop an open-source and open-data method for classification that is comparable in performance to paid services.
We then use the same data to audit the performance of our own and others' methods, shedding light on when classifiers fail, why they fail, and what, if anything, can be done in these instances to further improve performance.
Finally, we interrogate broader issues intrinsic to name-based gender classification, including how classification errors are unevenly distributed across groups. 

Throughout this work we use the labels female and male to refer to the gendered groups constructed by our classification scheme. This is intentional. There is a long history of thought on how best to theorize and talk about gender~\cite{mill1869subjection,de2010second,west1987doing,butler1988performative,scott1986gender,thorne1993gender,anderson1995knowledge,haslanger2000gender,mikkola2009gender,dembroff2018nonbinary,byrne2020women,dembroff2020escaping,hu2020sex} as well as more recent concern that name-based gender classification is flawed (due to how names are embedded in the gender binary) and potentially harmful (due to misgendering individuals) to such a degree that it ought to be used sparingly in scientific research, if at all~\cite{mullen2021gender, cheong2021computer}. 
Fully engaging with this literature and these concerns is beyond the scope of this paper. 
However, our language reflects our view, which is, in briefest form: names reify fictions about how one's social position is derivative of real (or perceived) sex-related characteristics. That is, names are a way of gendering, of projecting social life onto something thought to be more natural and thus definitive.
As such, the thoughtful exposure and use of name-gender associations can offer valuable opportunities for scientific and political work.
The somewhat strange locution that an individual is ``gendered female" or ``gendered male" used throughout is meant to capture the incoherence of the supposed ``sex/gender" dichotomy and to convey that naming or classifying is always an active process, a process as strange as the social phenomena our name-based gender classification scheme is designed to study.

\section{Methods}

\subsection{A Cultural Consensus Model of Name-Gender Associations}

The approach to name-based gender classification introduced in this section is motivated by three observations.
First, potential inaccuracies in source data could bias consensus estimates of how names are gendered.
Therefore, the model we use to combine source data and estimate a consensus should discover which sources are reliable and allow more reliable sources to more heavily influence the consensus.
Second, differences between data sources may not only be due to varying degrees of accuracy, but are potentially attributable to how a name's gendered association varies over time or between countries. A model of how names are gendered should be consistent with this possibility and allow for some means of specifying context to form more nuanced estimates.
Finally, how a name is gendered, even considering a single time and place, is a culturally constructed belief that emerges from individual views and actions, not a fixed feature of the natural world. This fact introduces a strong conceptual and practical constraint: our model must function in the absence of the kind of ground truth data used to train supervised meta-learning algorithms, such as stacking~\cite{wolpert1992stacked}.

To develop a name-based gender classification scheme consistent with the above observations, we leverage Cultural Consensus Theory (CCT), a method with origins in anthropology~\cite{romney1986culture}. CCT was first used to estimate a culture's consensus beliefs on a set of binary questions from individual sources' responses. Rather than merely averaging sources' responses, CCT uses a contextual estimate of each source's ``competence" to weight responses and better estimate the consensus. And, CCT uses each source's aggregate agreement with the current consensus to estimate that source's competence. Thus, CCT jointly infers both a culturally constructed consensus and each source's agreement with that consensus (i.e. their competence).

Here we adapt CCT to estimate how and how strongly each name is gendered, based on the cultural associations reflected in a corpus of reference data. Our twin goals are to (i) form a consensus estimate $y_m$ of how each name, indexed by $m$, is gendered, while also (ii) estimating the competence $c_n$ of each source, indexed by $n$. To do so, we directly incorporate the report $x_{n,m}$ of how a name $m$ is gendered by source $n$. Without loss of generality, we let $x_{n,m}$ take on a value of $1$ when source $n$ reports name $m$ as gendered female and $0$ when gendered male. CCT makes the mathematical assumption that a single cultural consensus exists, i.e. that $y_m$ is either $0$ or $1$, and infers a related quantity $z_m$, the probability that $y_m=1$, given the data and each source's competence.
In this way, $z_m$ provides a probabilistic measure of the strength of each name-gender cultural association, based on the available data; when names do not have a strongly gendered cultural consensus, $z_m$ will be closer to $\tfrac{1}{2}$.

The mathematical derivation of CCT, originally introduced in \cite{romney1986culture}, begins by using Bayes' theorem with an uninformative prior to write down the probability that name $m$ is gendered female, given the data sources and their competences,
\begin{equation}\begin{split}
    P(y_m = 1 \mid \mathbf{x}, \mathbf{c} ) \propto  
    \prod_{n=1}^{N} P(x_{n,m} \mid y_m =1, c_n)\ .
    \label{eq:formulation}
\end{split}\end{equation}

Note that, were the data sources ($\mathbf{x}$) or the competences of those sources ($\mathbf{c}$) to change, then the estimated consensus would also change. The next step in the derivation is to observe that $P(x_{n,m} \mid y_m=1, c_n)$ is the likelihood of observing data $x_{n,m}$, given a consensus $y_m=1$ and a competence $c_n$. By explicitly defining competence $c_n$ to be the probability that source $n$ correctly reports the consensus, independent of the name, we can write $P(x_{n,m} \mid y_m = 1, c_n) = x_{n,m} c_n + (1-x_{n,m})(1-c_n)$. In other words, if the consensus is $y_m=1$, the likelihood that a data source reports $1$ is $c_n$ and the likelihood it reports $0$ is $1-c_n$. Plugging this likelihood into Eq.~\eqref{eq:formulation}, and using the notation $z_m \equiv P(y_m=1 \mid \mathbf{x}, \mathbf{c})$, we get 
\begin{equation}\begin{split}
    z_m = \prod_{n=1}^{N} \left [ x_{n,m} c_n + (1-x_{n,m})(1-c_n) \right ] \Bigg/\\
    \bigg(\prod_{n=1}^{N} \left [ x_{n,m} c_n + (1-x_{n,m})(1-c_n) \right ] +\\ \prod_{n=1}^{N} \left [ x_{n,m} (1-c_n) + (1-x_{n,m})c_n \right ] \bigg)\ .
    \label{eq:expectation}
\end{split}
\end{equation}

While Eq.~\eqref{eq:expectation} relates our name-gender estimate $z_m$ to the data, it also depends explicitly on knowing the competences $\mathbf{c}$. In turn, the competence of each source depends on its responses $\mathbf{x}$ and the probability that the consensus is either $1$ ($z_m$) or $0$ ($1-z_m$) across all the queried names. This allows us to write the competence $c_n$ of source $n$ as that source's average agreement with the consensus $z_m$, 
\begin{equation}
    c_n = \frac{1}{M} \sum_{m=1}^{M} x_{n,m} z_m + (1-x_{n,m})(1-z_m)\ .
    \label{eq:maximization}
\end{equation}

Taken together, Eqs.~\eqref{eq:expectation} and \eqref{eq:maximization} recursively relate both the consensus and the competencies to each other, through the data. To fit this model to data (and thus estimate how each name is gendered), we use an Expectation Maximization (EM) algorithm: during the ``expectation'' step, name-gender associations $\mathbf{z}$ are updated using the current estimates of each source's competence $\mathbf{c}$~(Eq.~\eqref{eq:expectation}). Then, during the ``maximization'' step, source competences $\mathbf{c}$ are updated using the current estimates of how names are gendered $\mathbf{z}$~(Eq.~\eqref{eq:maximization}). These two steps are repeated until convergence, when neither $\mathbf{z}$ nor $\mathbf{c}$ changes, and the algorithm exits. Initially, all competences are assumed to be 0.9, but algorithm outputs do not depend on a particular choice of this value, provided that it is greater than 0.5. In this fashion our CCT approach provides reliable estimates of both consensus name-gender associations and source quality, without reference to any ``ground truth" data. An implementation of this CCT model and EM algorithm can be found with the code accompanying this paper \cite{vanbuskirk2022gender}.

\subsection{Gendered Name Data}

The CCT model for cultural name-based gender classification estimates its consensus directly from data. As a consequence, the broader the data on which it relies, the broader the consensus it estimates, and the larger the set of names for which it can compute a consensus at all. To support these needs, we collected and organized data spanning over 150 countries and more than a century from 36 publicly available sources with the goal of assembling a broad and globally representative dataset of names and their gendered associations. Sources include the U.S.\ Social Security Administration's record of baby names, a database of Olympic athletes, and WikiData, among others (See Table~\ref{tab:Sim1}). In total our compiled reference dataset captures gendered associations as well as global and temporal changes in naming practices for over 500,000 unique names.

Cleaned, curated versions of each source's data are available alongside source data in its raw form \cite{vanbuskirk2022osf}.
Making data open source in these ways not only provides researchers essential contextual information but, importantly, because our CCT classifier infers a consensus over the sources provided, researchers may broaden or refine the consensus by further expanding or restricting the data used.
In contrast to paid services, our organization and methodology encourages, rather than hinders, this kind of adjustment and extension via the incorporation of newly found sources or updating of an existing source.

In addition to the above we provide an open source Python package for researchers to conduct name-based gender classification in scientific contexts \cite{vanbuskirk2022nomquamgender}.
We make two important kinds of adjustments to source data in building this tool.
First, we drop diacritics, clip multipart names, and romanize where relevant. We do the same processing for names in the sample study to enable smooth and consistent matching between query and reference data.
Second, we attempt to account for the oversampling of either males or females that we see in some sources (e.g. Wikidata and the database of Olympic athletes).
By upweighting the underrepresented group (a kind of post-stratification) we ensure our classifications do not perpetuate a known bias present in the empirical data.
Together, the resources we provide allow researchers to deeply engage with name-gender association data or quickly make classifications depending on the needs of their specific research project.

\subsection{Validation Data}

We evaluate our CCT classifier using three validation datasets. 
The first is a dataset of 5,778 academic authors compiled by Santamaría and Mihaljevíc circa 2018~\cite{santamaria2018comparison}, curated from various bibliometric studies as a means by which to evaluate ``name-to-gender inference" services. The majority of this dataset's ``human-annotated author-gender data" came from inspecting online personal information or photographs associated with each author.
The second is a dataset compiled circa 2012 from the study of 10,464 researchers in the field of Natural Language Processing~\cite{vogel2012he}. The authors of this study ``labeled the gender of authors\ldots mostly manually" via the use of personal cognizance, photos, and name lists. 
The third is a dataset comprising 7,188 U.S.\ academics whose responses to the question ``What is your gender?" were collected as part of a 2021 survey of academic faculty~\cite{morgan2021socioeconomic}. The vast majority of respondents identified as either ``Female" or ``Male", while those who selected ``Other identity" or ``Prefer not to say" made up less than 1\% of respondents and are not included in this validation dataset. The Santamaría and Vogel datasets are included among the 36 public sources; the survey data collected by Morgan et al.\ is not publicly available due to IRB agreements. To avoid validating our algorithm on data used in training, we employ the standard practice of splitting our data into disjoint sets for training and subsequent validation.

\subsection{A Taxonomy of Names}

\begin{figure}[ht]
      \centering
      \includegraphics[width=1.0\linewidth]{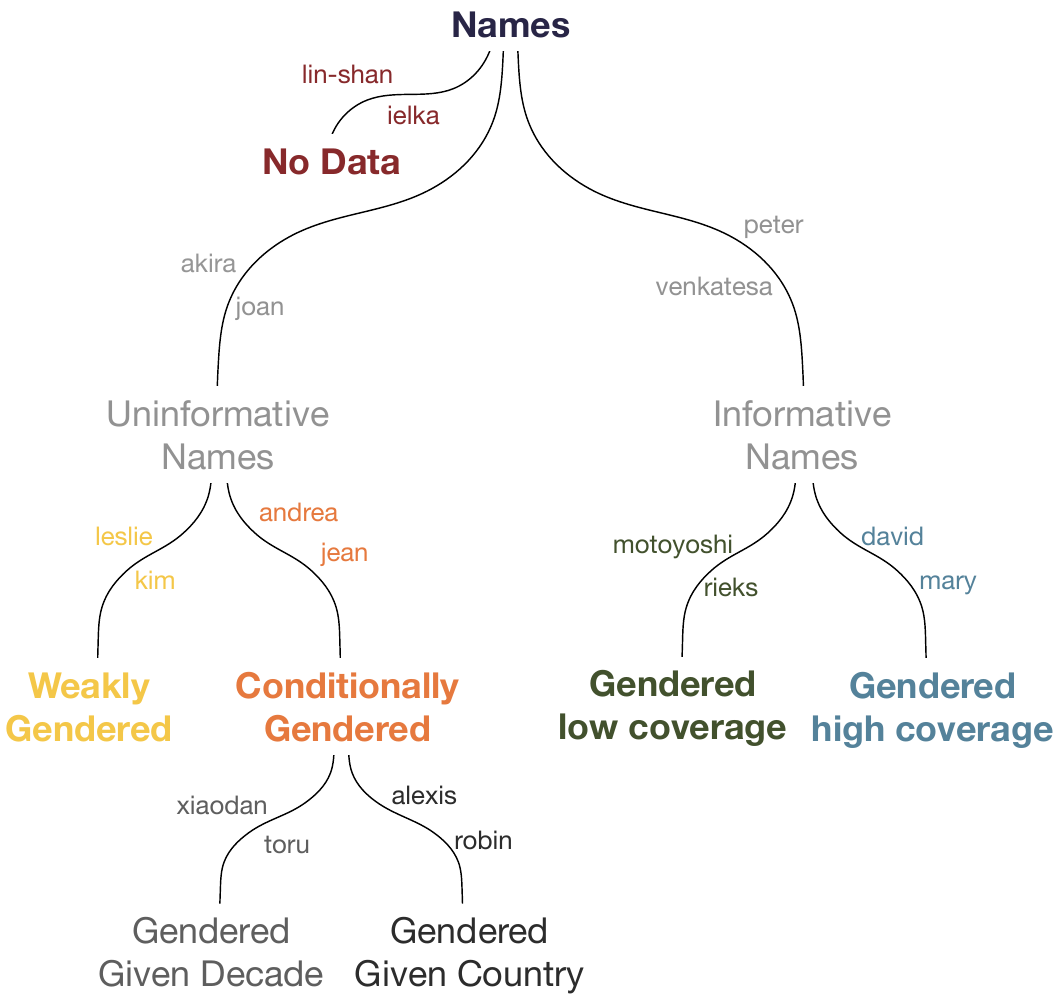}
      \caption{A taxonomy constructed to hierarchically classify each name in a sample based on how much data we have on the name (e.g. no data, low coverage), how strongly the name is gendered (e.g. uninformative, informative), and whether or not the properties of the name change over time or across countries (e.g. weakly gendered, conditionally gendered).}
      \label{fig:taxonomy}
\end{figure}

To complement the algorithm and data of this study, we also introduce a taxonomy of names, designed with name-based gender classification in mind (Fig.~\ref{fig:taxonomy}). This taxonomy divides the names themselves into categories, based on empirically measurable properties including how common the name is, how strongly it is gendered, and whether or not these properties vary with temporal or cultural context.
This taxonomy therefore has dual goals, enabling (i) an understanding of the idiosyncratic makeup and challenges of particular names (and therefore sets of names), and (ii) an ability to diagnose the quality of name-based gender classification algorithms by evaluating their performance across each leaf of the taxonomy.

To categorize a name, our proposed taxonomy draws on the 36 datasets described above, or any extension or refinement thereof, which we hereafter refer to simply as reference data. First, if a name is not observed in the reference data, we set it aside. Otherwise, we evaluate whether a name is informative according to the reference data: a name is informative if we have low uncertainty about the gendered group to which we ought to assign individuals with that name. We formalize informativeness and uncertainty by computing the entropy of the empirical distribution over gendered groups for each name in the reference data. We define an informative name as having an entropy of at most 0.47 bits, equivalent to requiring that our reference data suggests there is a probability greater than or equal to 0.9 that someone with the name ought to be classified as either gendered male or gendered female.

Informative names are further divided into those with low and high coverage. Here, high coverage names are those observed in our reference data at least $m$ times, with $m=10$. In our reference data, for instance, Alexander is observed $\sim10^6$ times, Madelyn $\sim10^5$, Abdellah $\sim10^4$, Garreth $\sim10^3$ and Finneus $\sim10^2$. 
Uninformative names are divided into two categories, motivated by the observation that a name may be uninformative because the consensus name-gender association is either (i) weak in general, or (ii) strong, but varying over time~\cite{blevins2015jane, smith2013search} or across countries~\cite{vogel2012he, torvik2016ethnea}. We formalize this difference by computing the conditional entropy of the name, conditioned on country or decade of origin. If both the country-conditional entropy and the decade-conditional entropy remain above 0.47 bits, the name is classified as weakly gendered. However, if either of the conditional entropies is below 0.47 bits, the name is classified as conditionally gendered--either gendered given decade or gendered given country. Names falling into both categories of conditionally gendered names are classified as gendered given country. Names that would be conditionally gendered given both country and decade, but not given country alone or given decade alone, are categorized as weakly gendered.

As a technical note, the computation of a conditional entropy takes the form of a weighted average, in which the weights require the incorporation of the probability of observing each piece of conditional data. Here, such calculations must incorporate the probability of a particular country or decade, given a name: $P(\mathbf{c} \mid \text{name})$. To calculate this probability, we employ Bayes' rule, $P(\mathbf{c} \mid \text{name}) \propto P(\text{name} \mid \mathbf{c})\ P(\mathbf{c})$. While $P(\text{name} \mid \mathbf{c})$ can be computed directly from source data, $P(\mathbf{c})$ represents a prior distribution over the origins of names more broadly, requiring one to make a choice between an uninformative prior, $P(\mathbf{c})=\text{const}$, which aims to represent no {\it a priori} belief about country of origin, and an informative prior, $P(\mathbf{c})$, which seeks to reflect the relative representation of countries in a sample.
To construct an informative prior, one could potentially use global data about population sizes over geographies and history, or one could alternatively incorporate beliefs about the origins of the study population at hand. The results that follow remain agnostic, however, and all taxonomic categorizations use an uninformative prior.

\subsection{Comparisons to Paid Services}

To further contextualize our method's performance, we compare the classifications our method makes for the individuals in the Santamaría 2018 dataset with those of four paid services:  NamSor, Gender API, Onograph, and genderize.io. NamSor, Gender API, and genderize.io were previously evaluated on this dataset and we make use of the classifications cached from that analysis~\cite{santamaria2018comparison}. In addition, we queried these three services in February, 2021, for those individuals which were not classified when queried in 2018. This resulted in further classifications for 1,016 instances of a total of 1,252 previously unclassified individuals. We are unaware of any published evaluation of the service Onograph and queried it for all individuals in the Santamaría 2018 dataset.

\section{Results}

The analyses that follow utilize our proposed taxonomy of names to demonstrate three key findings. We first show that the vast majority of names in our validation datasets are common and strongly gendered, and that the taxonomic composition of datasets varies only somewhat. We then show that variation in classifier performance is largely explained by the composition of the dataset in question---in particular its proportions of uninformative and unclassified names. In this context, we show finally that more nuanced treatment of those most challenging names provides only marginal improvement, due to existing methods being already well calibrated to empirical data.

\subsection{A Taxonomy-informed Assessment of Datasets}

\begin{figure}[t]
  \centering
  \begin{subfigure}{0.495\columnwidth}
  \includegraphics[width=\textwidth]{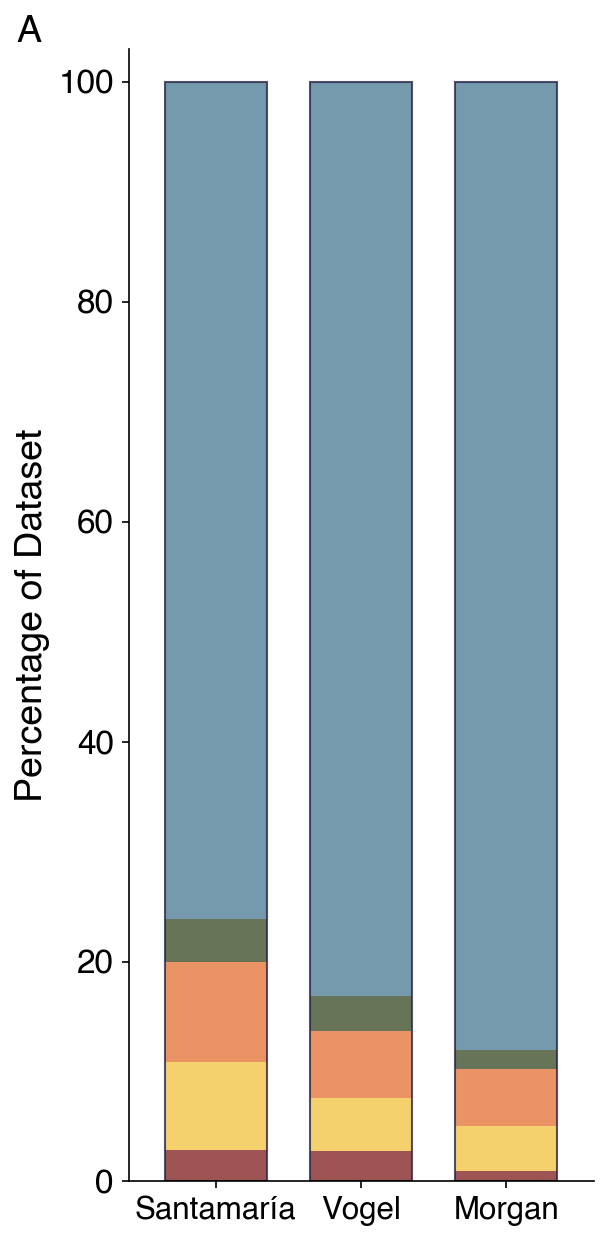}
  \label{fig:composition}
  \end{subfigure}
  \hspace{-1.1em}
  \begin{subfigure}{0.505\columnwidth}
  \includegraphics[width=\textwidth]{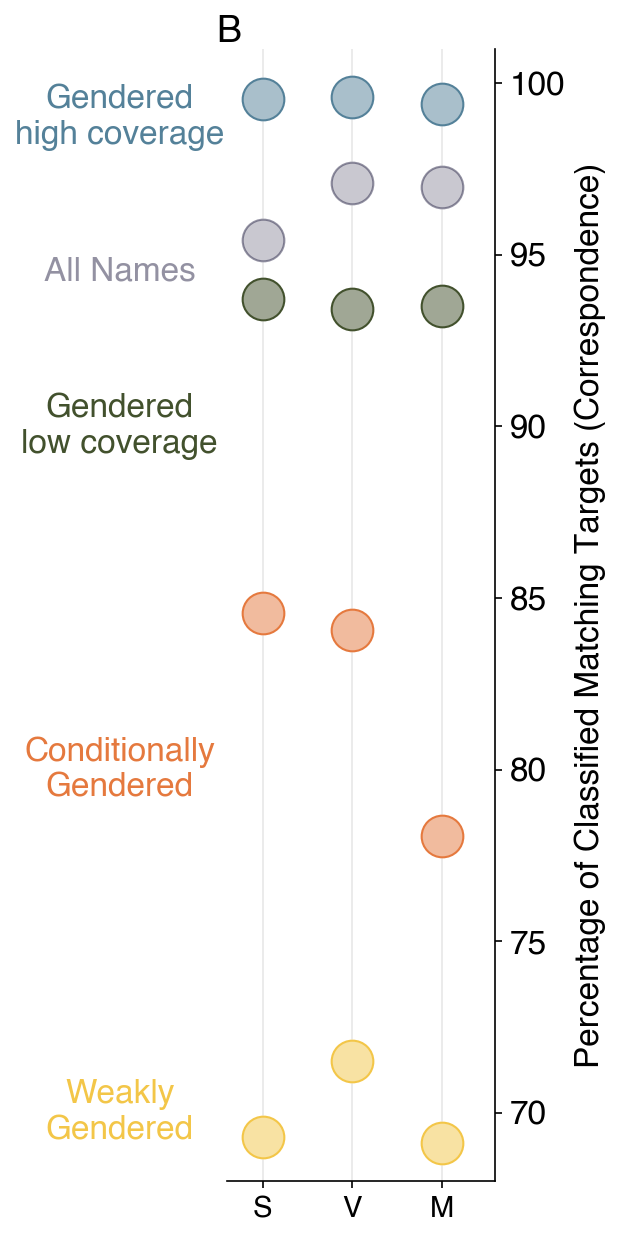}
  \label{fig:datasets}
  \end{subfigure} 
      \caption{Understanding sample composition and classifier performance through the lens of a taxonomy of names. (A) Percentage of individuals in each of the three datasets used for validation that have names in each of five exclusive and comprehensive leaves of the taxonomy. (B) Percentage of individuals classified by our CCT Classifier assigned a class that matches given labels, broken out by taxonomic leaf.}\label{fig:overview}
\end{figure}

Common patterns emerge when different datasets are sorted into the leaves of our taxonomy (Fig.~\ref{fig:taxonomy}). First, the vast majority of individuals' names are gendered with high coverage. These individuals, with names such as Anita, Jose, and Fatima, make up 76.2\%, 83.2\%, and 88.0\% of the Santamaría, Vogel, and Morgan validation datasets, respectively (Fig.~\ref{fig:overview}A). As a consequence, most methods for name-based gender classification will perform well for most people, because most people have names with strong name-gender cultural associations.

The second largest group across all datasets is made of those individuals with conditionally gendered names, such as Andrea, Toru, and Alexis. Making up 9.1\%, 6.0\%, and 5.2\% of the samples, if additional data, such as year of birth or country of origin, were available one could potentially improve performance on these names by a meaningful margin, and by extension improve overall performance. We explore this possibility later.

Those with weakly gendered names, such as Akira, Leslie, and Kim, constitute 8.0\%, 4.8\%, and 4.0\% of individuals of the Santamaría, Vogel, and Morgan validation datasets, respectively (Fig.~\ref{fig:overview}A). This means that regardless of methodology there will remain a nontrivial portion of individuals with names that resist efforts to construct gendered groups. 
In other words, all methods must confront the fact that some names are simply not strongly gendered, and individuals with these names will therefore be classified incorrectly at the Bayes error rate---a predicted 33\% of the time for Akiras and Kims, 42\% of the time for Leslies.

While we have emphasized similarities in the taxonomic composition of these datasets, it is also true that there are twice as many individuals with weakly gendered names in the Santamaría dataset than in the Morgan dataset (8.0\% vs 4.0\%), and almost 4\% more individuals with conditionally gendered names (9.1\% vs 5.2\%). As a result, in the absence of additional information on which to condition classifications, the Santamaría dataset presents all potential name-based gender classification methods with a larger set of individuals for whom the error rate is expected to be substantial. Indeed, this expectation is borne out unambiguously in our subsequent analyses.

\subsection{A Taxonomy-informed Assessment of Classifiers}

Through the lens of our taxonomy, the performance of any method will depend on two factors: the relative proportions of those categories in the data under investigation---explored in the previous section---and the method's performance in each of those categories. In other words, overall performance can be decomposed into a weighted average of performances across taxonomic categories. Here, we confirm that the performance of our CCT classifier is greatly determined by the taxonomic categories of the names classified. Thus, while our method's classifications match given labels around $96$\% of the time, we next show that this good overall performance, and variation therein, are driven by the combined consequences of dataset composition and per-category performance.

\begin{figure*}[t]
      \captionsetup[sub]{labelformat=empty,aboveskip=-100pt}
      \centering
      \resizebox{1.\textwidth}{!}{%
      \subcaptionbox{\label{fig:services}}
        {\includegraphics[height=3cm]{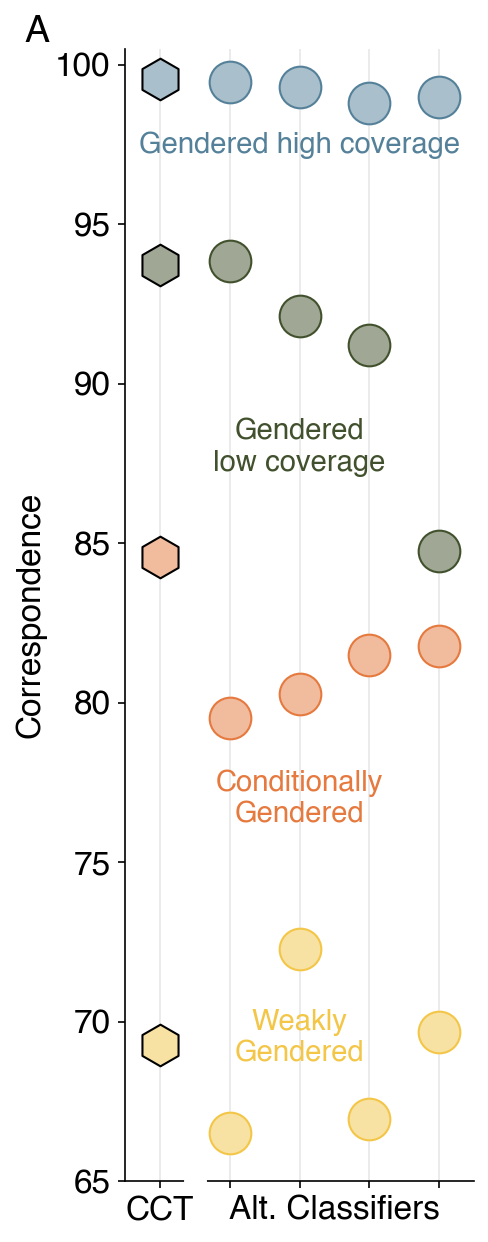}}
      
      \subcaptionbox{\label{fig:classified}}
        {\includegraphics[height=3cm]{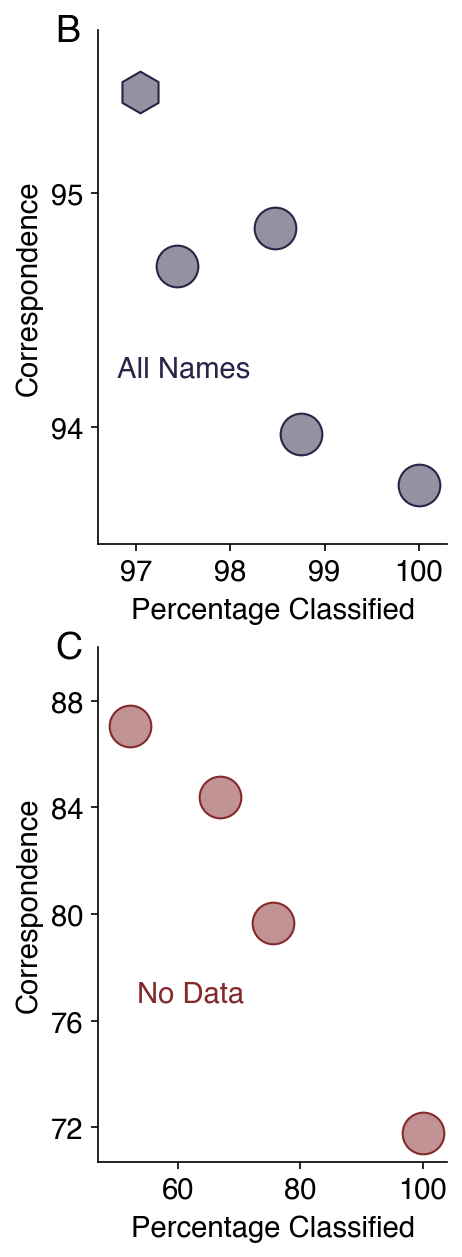}}
      
      \subcaptionbox{\label{fig:bins}}
        {\includegraphics[height=3cm]{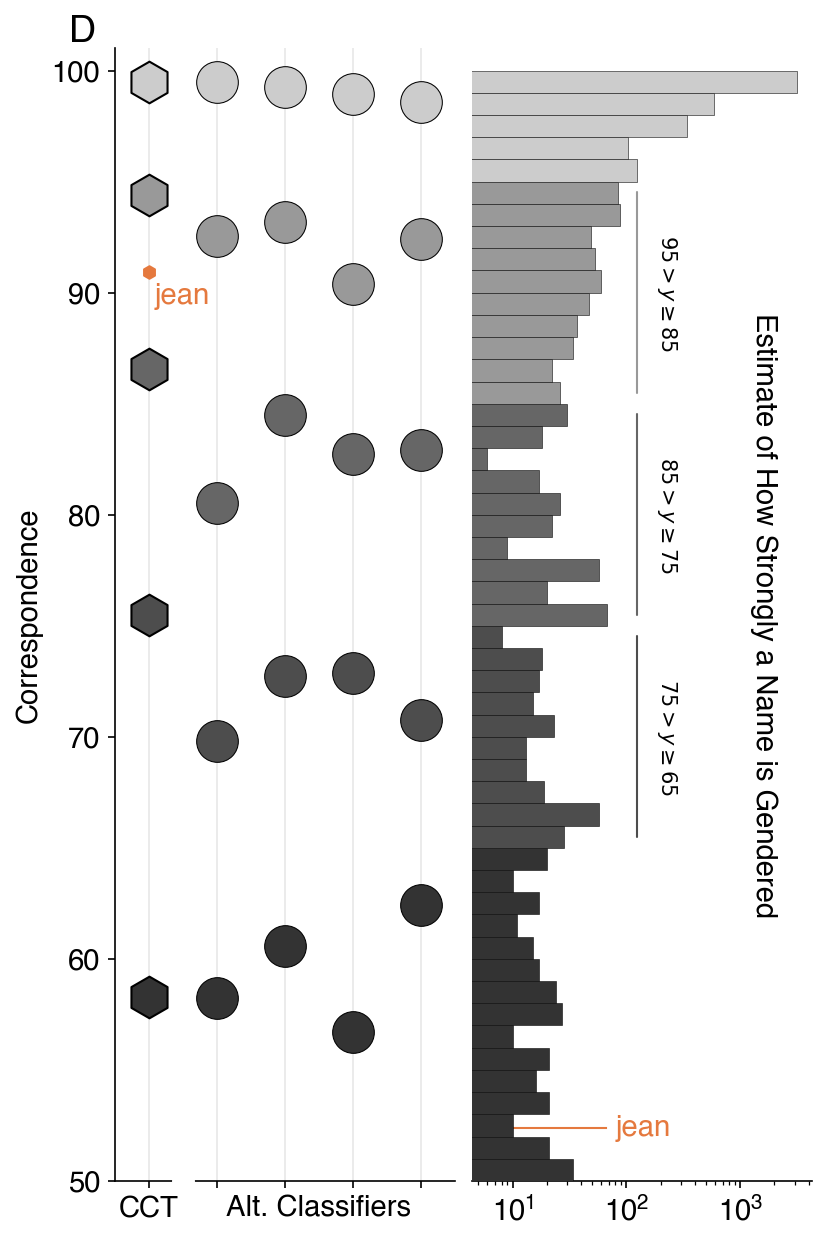}}
      }
      \caption{Evaluating performance of our CCT classifier and four paid services on the Santamaría 2018 validation dataset. (A) Percentage of classifications made by each classifier which match given labels, broken out by taxonomic leaf. (B, C) How performance correlates with the percentage of individuals a classifier classifies, the variation of which is greatly shaped by the number of individuals with names for which a classifier does not have data. (D) Performance of each classifier on subsets of individuals grouped based on how strongly their names are gendered. Those with the name Jean are held out of the subset to which they would be assigned.}\label{fig:santamaria}
\end{figure*}

First, across the three validation datasets, CCT's correspondence (the fraction of classifications made that match target labels) is consistently highest for names gendered with high coverage (${\sim}99$\%), moderate for names gendered with low coverage (${\sim}93$\%), and poor for weakly gendered names (${\sim}70$\%) (Fig.~\ref{fig:overview}B). In contrast, correspondence varied for conditionally gendered names, ranging from 85\% on the Santamaría dataset to 78\% on the Morgan dataset---a 7 percentage point difference. While names in this category are gendered when estimates are conditioned on a particular decade and/or country, here our CCT classifier provides unconditioned classifications. As a consequence, our classifier (and, by extension, any unconditioned classifier) will perform well for such names when a particular dataset aligns well with the global consensus (e.g., Santamaría), and poorly when it does not (e.g., Morgan). To illustrate this point, we note that the common conditionally gendered names Jean and Robin are gendered male in the global consensus, CCT classifications, and in the Santamaría dataset, but not in the Morgan dataset. If these two names are removed from consideration, the difference in CCT's correspondence on conditionally gendered names between the Santamaría and Morgan datasets drops from 7 percentage points to just 3. In short, correspondence was consistent across datasets for three taxonomic categories, and inconsistent for the fourth category which was context dependent by design.

Second, the coverage of our method---the proportion of names for which any classification is made at all---also affects overall performance. Our CCT classifier does not classify 3.0\%, 3.0\%, and 1.1\% of the three datasets respectively, almost always because the names in question are absent from the reference data (No Data; Fig.~\ref{fig:taxonomy}), and in rare instances because CCT's estimates were not numerically different from 0.5. In fact, variation in the proportions of weakly gendered and no-data names explains 80\% of variation in overall performance.

Taxonomic categories also explain the performance of other classifiers beyond our own. To demonstrate this, we evaluated four paid services (see Methods) on the Santamaría validation dataset and found that all services perform similarly with overall correspondence ranging from 95.4\% to 93.8\% (Fig.~\ref{fig:santamaria}B). 
As with our CCT classifier, the degree of correspondence with validation labels remained well characterized by taxonomic category, with all five methods producing non-overlapping taxonomic bands of performance (Fig.~\ref{fig:santamaria}A). This broad consistency across methods suggests that the reliability of classifications may be predicted in advance, based solely on how a target dataset is distributed across the leaves of our taxonomy.

Across methods, we observed two clear inverse correlations that help explain variation in performance. First, methods attempting more classifications decreased in overall correspondence, ranging from CCT classifying only 97.0\% of names and achieving 95.4\% correspondence, to Namsor classifying 100\% of names and achieving only 93.8\% correspondence (Fig.~\ref{fig:santamaria}B). Second, methods attempting more classifications of names falling into the no-data category decreased in correspondence for those names, ranging from genderize.io classifying only 52.1\% of names and achievieng 87.1\% correspondence to Namsor classifying 100\% of names and achieving 71.8\% correspondence. By definition, these names are rare, occurring zero times in the large corpus we assembled. Thus, marginal improvements in coverage---the set of names for which a method will return any classification whatsoever---come at the cost of lower correspondence with validation labels. 

\subsection{Model Calibration and Limits to Classification}

Our findings so far suggest that methods may be generally well calibrated, performing near the Bayes error rate for most names. To more directly show the calibration of the methods examined, we stratified classifier correspondence on the Santamaría data into five bands by the degree to which names are gendered. In the absence of ground truth, these degrees were measured by our reference data. No method does substantially better or worse than the reference data suggest is possible (Fig.~\ref{fig:santamaria}D). 

Apparent deviations from calibration reflect the idiosyncratic composition of the validation data. For example, our method, and all others, provide classifications for individuals named Jean matching approximately 90\% of validation labels, a percentage suggesting that Jean might be a moderately to strongly gendered name. However, our global corpus places Jean amongst the most weakly gendered names (Fig.~\ref{fig:santamaria}D). This discrepancy from calibration is explained by recognizing that Jean is a conditionally gendered name, exhibiting variation in usage across countries.
Figure~\ref{fig:santamaria}D is therefore presented with Jean separated out of the least gendered band, illustrating the risks of validation on small, idiosyncratic samples and reinforcing the broader trend of calibration.

\subsection{Conditioning Estimates on Country or Decade Data}
\label{sub:country}

Names like Jean, and the idiosyncrasies of the Santamaría dataset, suggest an appealing path to improved performance by using additional information to correctly classify names in the conditionally gendered taxonomic category. For instance, could performance be improved simply by knowing that most Jeans in the Santamaría dataset are French (and thus, more likely gendered male) rather than American (and thus, more likely gendered female)?  Torvik et al.\ explored such an idea with a paired gender and ethnicity classification method, ``Genni + Ethnea", which makes use of first and last names to condition gender classifications on ethnicity~\cite{torvik2016ethnea}. However, the extent to which this sophistication matters in practice has remained understudied. 

\begin{figure}[ht]
	\centering
	\includegraphics[width=1.0\linewidth]{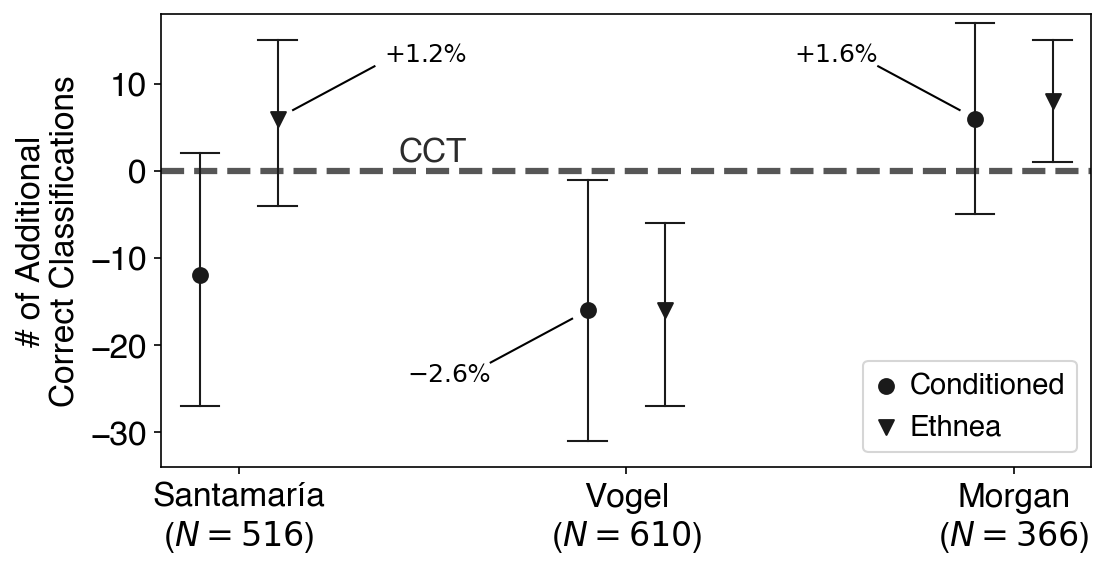}
	\caption{Comparison of our (unconditioned) CCT classifier and classifiers that use ethnicity data to localize classifications on the subset of names our taxonomy suggests are gendered given country. The change in the number of classifications that match given labels when conditioned classifications are used in place of CCT shows that conditioning does not meaningfully improve performance. The observed differences in the change in matching classifications is shown along with 95\% bootstrapped confidence intervals.}
	\label{fig:conditioning}
\end{figure}

To directly explore the value of conditioning our CCT method's classifications, we reprocessed names that are conditionally gendered given country data, which make up almost all of the conditionally gendered names in our three validation datasets (98\%, 97\%, 97\%). We first collected a predicted ethnicity from Ethnea for each first-last name pair~\cite{torvik2016ethnea}, and then converted each ethnicity to a country or set of  countries. Finally, we recomputed classifications for those names using only the country-specific reference data for each name, which we compared to the classifications made by the global consensus. Ethnea also returns a ``gender prediction" in some cases (via Genni), which we retained for additional comparison.

Perhaps surprisingly, conditioning on country data did not consistently improve performance on the names in the taxonomic category of conditionally gendered given country. Instead, bootstrapped confidence intervals of net change in performance show that there was no significant difference between the unconditioned CCT and country-conditioned CCT methods for the Santamaría and Morgan datasets, and significant decrease in performance for the Vogel dataset (Fig.~\ref{fig:conditioning}). We observed similarly mixed results when comparing to Ethnea-Genni directly, with no significant change in performance for the Santamaría dataset, significantly worse performance for the Vogel dataset, and significantly better performance for the Morgan dataset (Fig.~\ref{fig:conditioning}). These results may be driven by the fact that conditioning on the Ethnea-inferred country rarely leads to a different classification from that made by our CCT classifier: taking all three validation sets into consideration, Ethnea's inferences led to changes in classification in only 12\% of cases when country-conditioned CCT estimates were used, and in only 8\% of cases when using Ethnea-Genni directly. These low percentages place an upper bound on the potential impact---positive or negative---of attempts to improve name-based gender classification via conditioning on country.

\subsection{Attempting to Classify ``No Data" Names}

\begin{figure}[t]
	\centering
	\includegraphics[width=1.0\linewidth]{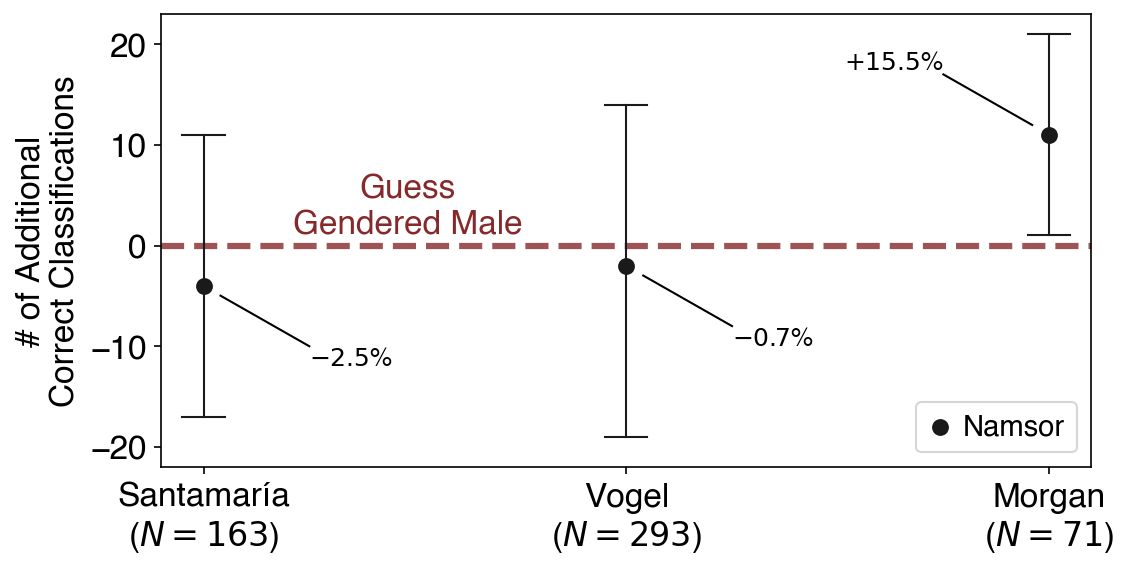}
	\caption{Comparison of Namsor with a simple ``Guess Male" heuristic on the subset of names for which we have no data. The change in the number of classifications that match given labels when Namsor is used in place of guessing male shows that Namsor does not meaningfully improve performance. The observed differences in the change in matching classifications is shown along with 95\% bootstrapped confidence intervals.}
	\label{fig:namsor}
\end{figure}

Our observations so far suggest that the classifications made by our method and others are unlikely to be improved substantially by further calibration (Fig.~\ref{fig:santamaria}D) or by conditioning estimates (Fig.~\ref{fig:conditioning}), leaving open only one further potential area of improvement: accurate classification in cases where no classifications are currently made. For our method, this would mean classifying individuals with names for which we have no data, i.e., names falling into the No Data taxonomic category. While collecting empirical data on these names is the obvious approach to enabling such classifications, we know of no additional, distinct, diverse, publicly available datasets not already included in our corpus. As an alternative, we therefore investigated the value of submitting all no-data names to Namsor, the paid service that returns classifications on 100\% of names. 

Among the no-data names of the Santamaría, Vogel, and Morgan validation datasets, Namsor's classifications matched validation labels 71.8\%, 76.8\%, and 76.1\% of the time. Namsor's average confidence measure for no-data names was only 0.36, compared to 0.9 for all other names. Together, these indicate that rare names are, indeed, more difficult to classify correctly. Nevertheless, the percentages of correspondence with validation labels above suggest that, perhaps, sending no-data names to Namsor would be a net improvement for overall accuracy. After all, one might argue that providing some classification is superior to providing none, as long as that classification is superior to simply guessing.

We now throw cold water on this line of thinking by showing through experiment that, unfortunately, using Namsor to classify no-data names is comparable to simply guessing. For the same 163, 293, and 71 no-data names in the Santamaría, Vogel, and Morgan datasets, respectively, we compared Namsor's performance in label correspondence to a guess-majority-class heuristic; for these three datasets, the majority class was gendered male. Namsor's performance was worse than simply guessing male for the Santamaría and Vogel datasets, and better for the Morgan dataset, with bootstrapped 95\% confidence intervals finding the former two differences not significant and the latter marginally significant (Fig.~\ref{fig:namsor}). In short, because the paid service performs similarly to guessing the majority class, we hesitate to recommend the use of a paid service to process the no-data taxonomic category.

\subsection{Averaging and Cultural Consensus Theory}

\begin{figure}[t]
	\centering
	\includegraphics[width=1.0\linewidth]{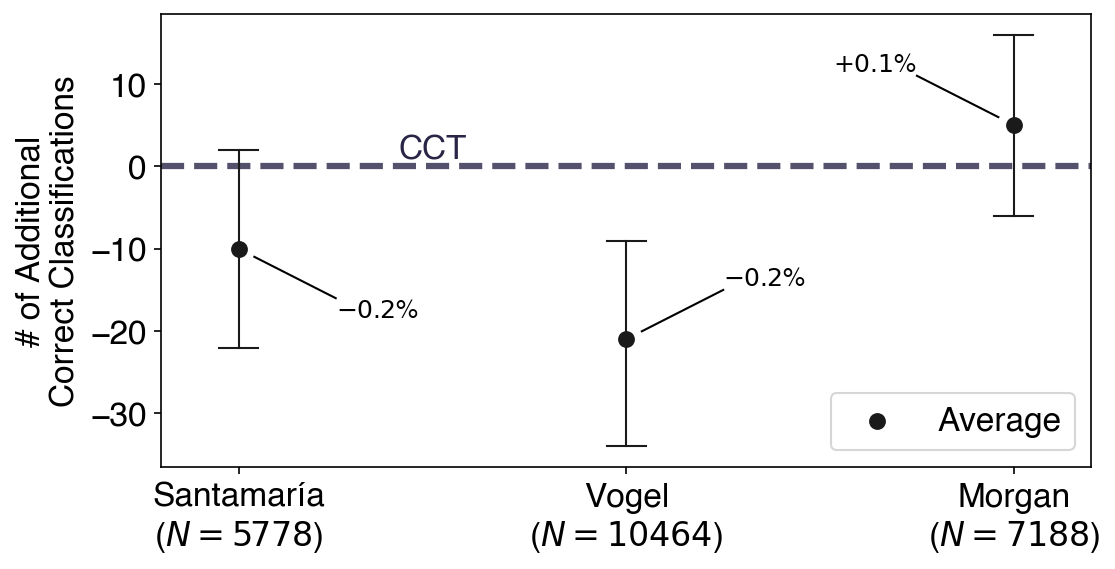}
	\caption{Comparison of performance of our CCT classifier and a simple averaging across sources. The change in the number of classifications that match given labels when averaging is used in place of CCT shows that CCT does not meaningfully improve performance. The observed differences in the change in matching classifications is shown along with 95\% bootstrapped confidence intervals.}
	\label{fig:avg}
\end{figure}

The results above suggest that all four paid services and our CCT classifier are well calibrated, and that additional improvements for conditionally gendered and no-data names will be difficult without additional data beyond first and last names alone. Together, these suggest that the empirical data we have collected enable one to achieve close to what appears to be the upper limit of what is possible with name-based gender classification. 
This observation led us to ask whether a simpler method that makes even more direct use of the empirical data could offer similar performance to that provided by our computationally sophisticated CCT classifier. 

To explore this possibility, we compared the performance of CCT, a sophisticated model of consensus and source competence, to a simple average consensus, a basic model that presumes equal competence across all 36 reference sources. Again using bootstrapped 95\% confidence intervals to estimate uncertainty, we found that the performance of the simple average was indistinguishable from CCT for the Santamaría and Morgan datasets, and 0.2\% worse for the Vogel dataset, equivalent to matching the validation labels for 20 fewer individuals (Fig.~\ref{fig:avg}). 

Nevertheless, the simple average consensus remained superior to simply using the classifications of a single data source.
For example, if one were to exclusively use data from the Social Security Administration's record of popular baby names to classify individuals in the Santamaría validation dataset, only 82.1\% of individuals receive classifications, of which 93.1\% match given labels. In contrast, the simple average consensus classified 96.8\% of individuals with 95.5\% correspondence. From the perspective of our taxonomy, the simple average consensus achieved a correspondence of 99.3\% on those 80.0\% of individuals with names that are either gendered with high or low coverage.

These observations recontextualize the importance of Cultural Consensus Theory in this work. Rather than serve as a means by which to construct a binary classifier, CCT primarily provides a way to ensure that each data source is reliable (i.e., high competence), and that there is a robust consensus across sources. It was only after evaluating sources with CCT that we could confidently experiment with other methods of classification, such as averaging. Importantly, CCT allows for evaluation directly against name-gender association data on thousands of names, sidestepping the idiosyncrasies of small samples. Going forward, CCT can provide a compelling alternative to more traditional means of assessing a source's general quality. This functionality will be critical as new sources are added to our reference set: researchers will be able to estimate source quality independent of any specific sample and then make an informed choice to include that source in the average, make use of it in limited capacity, or forgo it altogether.

\section{Discussion}

Here, we considered the general problem of characterizing the utility and performance of name-based gender classification methods. We surveyed existing methods and evaluated the practical and theoretical implications of the sources and consequences of performance variability. 
We then provided an open-source and transparent approach to name-based gender classification, with performance comparable or superior to existing state-of-the-art paid services. In tandem, this work also introduces a data-driven taxonomy of names which sheds light on the strength and stability of name-gender cultural associations across their potentially changing contexts. Underlying both of these contributions is a curated dataset combining 36 publicly available resources. The classifier, taxonomy, and curated data are freely available to support research that may rely on name-based gender classification \cite{vanbuskirk2022gender}.

Our taxonomy-informed evaluation of methods revealed that the empirical properties of names in a dataset determine methods' performance. In practice, the vast majority of names researchers are likely to encounter are strongly gendered names for which we have sufficient reference data. Our classifier achieves around 99\% correspondence with validation labels on these strongly gendered names, with all four paid services exhibiting similar performance. 

For other categories of names---be they taxonomic categories or simply groups of names that are similarly strongly gendered---all methods perform similarly on names of the same category. As a result, differences between methods in overall performance are driven less by how methods differ when they do make classifications and more by how many classifications they are able to make; when a method fails to provide any classification at all, it cannot correspond to any validation label, by definition. Nevertheless, we found that our method is not meaningfully improved by using a paid service to classify those names for which our method fails to make a classification for lack of data. And, in spite of the appeal of combining first and last names to infer nationality, to be then used to provide conditional classification, we found that empirically there was no consistent improvement in overall performance. 

In short, we find that future advances in name-based gender classification are fundamentally limited by the Bayes error rate, and thus any meaningful advances are more likely to come from the incorporation of additional data to improve coverage, and from the open-sourcing of methods and datasets whenever possible to improve transparency.

Our observations that most names are strongly gendered and that methods are well calibrated led us to critically examine the value of our Cultural Consensus Theory (CCT) approach. Indeed, across the three validation datasets used in this study, CCT agreed with a simple average consensus of the 36 source datasets for over 99\% of individuals, and we observed no significant changes in performance when the methods did disagree. In turn, this means that the simple average consensus of public data is comparable in performance to paid services. Further, the numeric name-gender estimate that comes from averaging across sources is of great potential use in post-processing classifications and assessing classifier confidence. For these reasons, the simple average consensus is provided in a Python package accompanying this paper~\cite{vanbuskirk2022nomquamgender}.

Nevertheless, CCT provides a valuable mathematical parallel to the social construction of gender. Rather than falling into the category of supervised learning, in which presumed ground-truth labels are used to train a classifier, CCT recognizes that name-gender associations are not a subject for which a ground truth exists. Instead, names---particular sequences of characters and phonemes---have been culturally gendered to greater or lesser degrees. Thus, the appropriate mathematical approach is to measure the consensus, giving higher weight to data sources that more reliably align with that consensus. It follows that the classifications outputted by such a method are nothing more than the consensus estimates of how each name is gendered, as agreed upon by the available reference data. 

Name-based gender classification raises issues of fairness, only one of which we directly address in this study. We showed that, for those names that are absent from the reference data, referred to as no-data names, one can achieve up to 77.5\% correspondence to labels by simply guessing male, the majority class. But this naive approach to improving performance leads to maximal imbalances in misclassification rates by class (0\% for those gendered male, 100\% for those gendered female). This result should serve as a cautionary warning to methods lacking transparency that seek to maximize overall ``accuracy.'' Doing so may improve top-line performance to the detriment of fairness and systemic bias.

These observations place name-based gender classification in a familiar position with respect to the desiderata of fairness. On the one hand, we might add a desideratum of ``fairness" at the level of gendered groups (related to the idea of sufficiency in the fair machine learning literature), via equal misclassification rates across groups, to our primary aim of meaningfully classifying a large portion of a sample~\cite{barocas-hardt-narayanan}. On the other hand, we might consider the total number of individuals of each group misclassified, called (somewhat confusingly) the ``Gender Bias Error Rate"~\cite{santamaria2018comparison,wais2016gender}. These counts are significant by virtue of their relation to estimating the composition of a sample: composition estimates will be unbiased only when there are equal numbers of misclassifications across all groups. Unfortunately, these two notions of fairness are in tension, and cannot in general be achieved simultaneously unless either (i) the sizes of the two groups are the same or (ii) no misclassifications are made, criteria established in the fair machine learning literature~\cite{chouldechova2017fair} and paralleled by earlier work in epidemiology~\cite{diggle2011estimating}.

To build a more fair classifier and better estimate a sample's overall gender composition, one could try to select a subset of names that are almost always classified correctly. Our taxonomic category of gendered names with high coverage provides one such subset, and more general thresholding schemes could be used to construct others. Although this approach may sidestep issues stemming from misclassification, it introduces new issues related to nonclassifications. For example, in the Vogel validation dataset 85\% of those gendered male have names gendered with high coverage whereas only 78\% of those gendered female do. Thus, because a different percentage of individuals gendered male and gendered female have names that meet the inclusion criteria used, different percentages of each group will be removed from analysis. Due to how factors like gender and nationality interact with and determine features of names and representations in samples, one is unlikely to find a subset of names that is almost always classified correctly and consistently splits those gendered female and those gendered male into classified vs nonclassified subsets in equal proportions. Fortunately, in practice it seems we can get relatively good estimates of composition using all individuals classified: we estimated 73.2\% of individuals in the Vogel validation set were gendered male with a target 72.7\%. Here, if one uses only names gendered with high coverage the estimate rises to 74.5\%.

One important limitation of our data and method, which is shared by existing methods, is their poor performance on Romanized (Pinyin) Chinese names. When analyzing the Santamaría dataset, nearly half (47\%) of individuals with a predicted ethnicity of Chinese had names falling into the no-data or weakly gendered taxonomic categories. Only 5\% of individuals with a predicted ethnicity of ``English" had names in either of those categories. As a consequence, even as our method's misclassification rates on this validation set are similar by gender (4.0\% for those gendered male vs 5.2\% for those gendered female), they are markedly higher for Romanized Chinese names. We are unaware of any method that overcomes the broad loss of name-gender associations in the Romanization process. This suggests that studies of globally representative datasets are unlikely to achieve balanced gender misclassification rates across nationalities if the names in those datasets are fully Romanized.

We identify three areas of future work that may further improve the scientific study of gender. First, growing the open reference data on which name-based gender classifications are based will improve coverage while maintaining transparency. This applies to both the addition of names not yet in the reference data, as well as increasing the depth of coverage of names already observed. Second, use of tools like the taxonomy introduced here may allow researchers to critically evaluate the potential uncertainty inherent in the sets of names they wish to analyze. Knowing that a large number of names fall within the weakly or conditionally gendered categories, as in the Santamaría dataset, should lead to increased caution. In addition, the reference data we have collected can be used to extend this taxonomy, for example cataloging which names are potentially sensitive to Romanization and other kinds of processing. Finally, name-based gender classification itself is inherently limited by its reliance on names alone. Future work could both improve performance and better reflect individuals' identities by shifting away from names alone and toward pronoun, survey, or other self-identified information. This will require that we continue to interrogate both the suitability of name-based gender classification for specific applications as well as how it fits into the broader ecosystem of methodologies we have to understand and improve social systems.

\section{Acknowledgements}
We thank M. Hoefer, N. LaBerge, K. Spoon, H. Wapman, and S. Zhang for valuable feedback. Funding: this work was supported, in part, by Air Force Office of Scientific Research Award FA9550-19-1-0329 (all authors) and the National Science Foundation under the Alan T. Waterman Award, Grant No. SMA-2226343 (DBL). 

\section{Competing interests}
The authors declare that they have no competing interests. 

\section{Data and materials availability}
All data needed to evaluate the conclusions in the paper are present in the paper, the Supplementary Materials, and/or available online: Data sets, \cite{vanbuskirk2022osf}; Reproduction code, \cite{vanbuskirk2022gender}; Python package, \cite{vanbuskirk2022nomquamgender}.

\bibliography{bibliography}

\clearpage
\onecolumn

\beginsupplement
\section{Supplementary Material}

\vspace{4em}
\centering

\Rotatebox{90}{%
\begin{threeparttable}
    \tiny
    \caption{Data Sources} 
    \label{tab:Sim1} 
    \begin{tabularx}{1.15\textwidth}{| c l X c |}
        
        \toprule
        \multirow{2}{*}{\bfseries \#} & 
        \multirow{2}{*}{\bfseries Source Name} & 
        \multirow{2}{*}{\bfseries File} & 
        \multirow{2}{*}{\bfseries Retrieved} \\
        \cmidrule(lr){1-4}

 1 & United States Social Security Administration (SSA) & \href{https://www.ssa.gov/oact/babynames/limits.html}{National data (7Mb)} & 02/13/2021  \\ \hline
 
 2 &  France Institut national de la statistique et des études économiques (Insee) & 
 \href{https://www.insee.fr/fr/statistiques/2540004\#consulter}{Fichiers France hors Mayotte (csv, 1 Mo)} &
 02/13/2021  \\ \hline
 
 3 & South Australian Government Data Directory (Data.SA) & 
 \href{https://data.sa.gov.au/data/dataset/popular-baby-names/resource/534d13f2-237c-4448-a6a3-93c07b1bb614}{Most popular Baby Names (1944-2013)} &
 02/13/2021  \\ \hline
 
 4 & Ontario, Canada Data Catalogue & 
 \href{https://data.ontario.ca/dataset/ontario-top-baby-names-male}{Ontario top baby names (male)} & 
 02/13/2021  \\ \hline
 
 4 & Ontario, Canada Data Catalogue & 
 \href{https://data.ontario.ca/dataset/ontario-top-baby-names-female}{Ontario top baby names (female)} & 
 02/13/2021  \\ \hline
 
 5 & Ireland Central Statistics Office (CSO) & 
 \href{https://data.cso.ie/table/VSA60}{Girls Names in Ireland with 3 or More Occurrences} & 
 02/14/2021  \\ \hline
 
 5 & Ireland Central Statistics Office (CSO) & 
 \href{https://data.cso.ie/table/VSA50}{Boys Names in Ireland with 3 or More Occurrences} & 
 02/14/2021  \\ \hline
 
 6 & Norway Statistics & 
 \href{https://www.ssb.no/en/statbank/table/10467/}{10467: Born persons, by girls' name and boys' name 1880 - 2020} & 
 02/14/2021  \\ \hline
 
 7 & New Zealand Department of Internal Affairs (TTT) & 
 \href{https://www.dia.govt.nz/diawebsite.nsf/wpg_URL/Services-Births-Deaths-and-Marriages-Most-Popular-Male-and-Female-First-Names?OpenDocument}{Top 100 boys' and girls' names from 1954 to 2017 (.xlsx, 226KB)} & 
 02/13/2021  \\ \hline
 
 8 & Hungary Statistics & 
 \href{https://www.nyilvantarto.hu/hu/statisztikak}{First name statistics among those born in the previous year} & 
 02/13/2021  \\ \hline

 9 & Sweden Statistik Databasen (SCB) & 
 \href{https://www.statistikdatabasen.scb.se/pxweb/en/ssd/START__BE__BE0001__BE0001D/BE0001Nyfodda}{Newborns, first names normally used, by year of naming and sex. Year 1998 - 2020} & 
 02/13/2021  \\ \hline

 10 & Turkish İSTATİSTİK VERİ PORTALI (TUIK) & 
 \href{https://data.tuik.gov.tr/Search/Search?text=\%C4\%B0simleri}{Erkek İsimleri \& Kadın İsimleri} & 
 02/13/2021  \\ \hline

 11 & Austria Newborn Baby Names & 
 \href{http://www.statistik.at/web_de/statistiken/menschen_und_gesellschaft/bevoelkerung/geborene/vornamen/index.html}{List of all first names ever given 1984-2019} & 
 02/13/2021  \\ \hline

 12 & Belgium Statistics & 
 \href{https://statbel.fgov.be/nl/themas/bevolking/namen-en-voornamen/voornamen-van-meisjes-en-jongens\#panel-13}{Voornamen van meisjes en jongens} & 
 02/13/2021  \\ \hline

 13 & Switzerland Federal Statistical Office (FSO) & 
 \href{https://www.bfs.admin.ch/bfs/en/home/statistics/population/births-deaths/first-names-switzerland.assetdetail.13707161.html}{Female first names of the population by year of birth, Switzerland and language regions, 2019} & 
 02/13/2021  \\ \hline

 13 & Switzerland Federal Statistical Office (FSO) & 
 \href{https://www.bfs.admin.ch/bfs/en/home/statistics/population/births-deaths/first-names-switzerland.assetdetail.13707162.html}{Male first names of the population by year of birth, Switzerland and language regions, 2019} & 
 02/13/2021  \\ \hline

 14 & Scotland National Records (NRS) & 
 \href{https://www.nrscotland.gov.uk/statistics-and-data/statistics/statistics-by-theme/vital-events/names/full-lists-of-babies-first-names-archive}{Full Lists of Babies’ First Names - Archive} & 
 02/13/2021  \\ \hline

 15 & Northern Ireland Statistics and Research Agency (NISRA) & 
 \href{https://www.nisra.gov.uk/sites/nisra.gov.uk/files/publications/Full_Name_List_9718.xlsx}{Full Names List, 1997-2018 - TablesExcel (3.3 MB)} & 
 02/13/2021  \\ \hline

 16 & United Kingdom Office for National Statistics (ONS) & 
 \href{https://webarchive.nationalarchives.gov.uk/20201003225507/https://www.ons.gov.uk/peoplepopulationandcommunity/birthsdeathsandmarriages/livebirths/datasets/babynamesinenglandandwalesfrom1996}{Baby names in England and Wales: from 1996} & 
 02/13/2021  \\ \hline

 16 & United Kingdom Office for National Statistics (ONS) & 
 \href{https://webarchive.nationalarchives.gov.uk/20160108061941/http://www.ons.gov.uk/ons/publications/re-reference-tables.html?edition=tcm\%3A77-243767}{Baby Names, England and Wales, 1904-1994} & 
 02/13/2021  \\ \hline

 17 & Finland Open Data (Avoindata) & 
 \href{https://www.avoindata.fi/data/en_GB/dataset/none/resource/08c89936-a230-42e9-a9fc-288632e234f5}{Etunimitilasto 2021-02-05 DVV.xlsx} & 
 02/13/2021  \\ \hline

 18 & Slovenia Statistical Office (SiStat) & 
 \href{https://pxweb.stat.si/SiStatData/pxweb/en/Data/-/05X1010S.px}{Female first names, Slovenia, annually} & 
 02/13/2021  \\ \hline
 
 18 & Slovenia Statistical Office (SiStat) & 
 \href{https://pxweb.stat.si/SiStatData/pxweb/en/Data/-/05X1005S.px}{Male first names, Slovenia, annually} & 
 02/13/2021  \\ \hline

 19 & Spain Instituto Nacional de Estadística (INE) National & 
 \href{https://www.ine.es/dyngs/INEbase/es/operacion.htm?c=Estadistica_C\&cid=1254736177009\&menu=resultados\&secc=1254736195454\&idp=1254734710990\#!tabs-1254736195454}{All names with frequency equal to or greater than 20 people} & 
 02/13/2021  \\ \hline

 20 & Spain Instituto Nacional de Estadística (INE) International & 
 \href{https://www.ine.es/dyngs/INEbase/es/operacion.htm?c=Estadistica_C\&cid=1254736177009\&menu=resultados\&secc=1254736195454\&idp=1254734710990\#!tabs-1254736195454}{The 10 most frequent names of the most representative nationalities at the national level} & 
 02/13/2021  \\ \hline

 21 & Baby names in Japan, 2004–2018: common writings and their readings & 
 \href{https://osf.io/cdt6j/}{Data\_Baby\_Names\_Japan.xlsx} & 
 02/14/2021  \\ \hline

 22 & "gender" by Jörg MICHAEL & 
 \href{https://autohotkey.com/board/topic/20260-gender-verification-by-forename-cmd-line-tool-db/}{ftp://ftp.heise.de/pub/ct/listings/0717-182.zip} & 
 02/14/2021  \\ \hline

 23 & Facebook Generated Name List (FBNL) & 
 \href{https://sites.google.com/site/facebooknamelist/namelist}{firstname.csv} & 
 02/16/2021  \\ \hline

 24 & Kantrowitz Name Corpus & 
 \href{https://www.cs.cmu.edu/Groups/AI/util/areas/nlp/corpora/names/}{female.txt} & 
 02/16/2021  \\ \hline
 
 24 & Kantrowitz Name Corpus & 
 \href{https://www.cs.cmu.edu/Groups/AI/util/areas/nlp/corpora/names/}{male.txt} & 
 02/16/2021  \\ \hline

 25 & The Gender Gap in Science / Genderize.io & 
 \href{https://osf.io/7kbyf/}{pubmed\_gender\_data.sqlite3} & 
 02/16/2021  \\ \hline

 26 & Genni + Ethnea & 
 \href{https://databank.illinois.edu/datasets/IDB-9087546}{genni-ethnea-authority2009.tsv} & 
 02/16/2021  \\ \hline

 27 & 120 years of Olympic history: athletes and results & 
 \href{https://www.kaggle.com/heesoo37/120-years-of-olympic-history-athletes-and-results?select=athlete\_events.csv}{athlete\_events.csv} & 
 02/16/2021  \\ \hline

 28 & WikiProject Names & 
 \href{https://m.wikidata.org/wiki/Wikidata:WikiProject_Names/lists/given_names}{Wikidata:WikiProject Names/lists/given names} & 
 02/17/2021  \\ \hline

 29 & Behind the Name & 
 \href{https://www.behindthename.com/names/list.php}{Names by Usage} & 
 02/17/2021  \\ \hline

 30 & Wikidata Human Entities & 
 \href{https://query.wikidata.org/}{Wikidata Query Service} & 
 02/17/2021  \\ \hline

 31 & Viaf Personal Entities & 
 \href{http://viaf.org/viaf/data/}{viaf-20210201-clusters.xml.gz} & 
 02/19/2021  \\ \hline

 32 & Gender in the ACL Anthology Name lists & 
 \href{https://nlp.stanford.edu/projects/gender.shtml}{acl-male.txt \& acl-female.txt} & 
 04/26/2021  \\ \hline

 33 & Geoff Gender Guesser & 
 \href{https://www.gpeters.com/names/baby-names.php}{Most Masculine Names \& Most Feminine Names} & 
 04/26/2021  \\ \hline

 34 & Chinese-Tool Names & 
 \href{https://www.chinese-tools.com/names}{Girl names in Chinese \& Boy names in Chinese} & 
 04/26/2021  \\ \hline

 35 & Clauset et al. 2015 Census & 
 \href{https://www.science.org/doi/10.1126/sciadv.1400005}{All Faculty Name Gender Dictionary} & 
 08/30/2019  \\ \hline
 
 36 & Comparison and benchmark of name-to-gender inference services & 
 \href{https://github.com/GenderGapSTEM-PublicationAnalysis/name_gender_inference}{name\_gender\_inference/test\_data/raw\_data/all.csv} & 
 04/26/2021 \\
    
        \bottomrule
    \end{tabularx}
    \begin{tablenotes}
        \small
        \item More information along with cleaned, downloadable versions of these data sources can be found  \href{https://osf.io/tz38q}{online}.
    \end{tablenotes}
\end{threeparttable}
}

\end{document}